\documentclass[aps,prB,amsmath,amssymb,superscriptaddress,reprint,twocolumn]{revtex4-1}

\usepackage{graphicx}
\usepackage{setspace}
\usepackage{color}
\usepackage{natbib}
\usepackage{multirow}
\usepackage{bm}  % bold math
\usepackage{amsfonts,amsmath}
\usepackage{amssymb}
\usepackage{enumerate,ulem}
\usepackage{float}
\usepackage{ulem} % allow the strikethrough command \sout
\usepackage{epsfig,graphicx,color,calc,cancel}

\definecolor{darkgreen}{RGB}{32,127,16}

\newcommand{\beq}{\begin{equation}}
\newcommand{\eeq}{\end{equation}}
\newcommand{\ba}{\begin{align}}
\newcommand{\ea}{\end{align}}
\renewcommand{\phi}{\varphi}

   \let\th\theta  \let\b\beta \let\a\alpha \let\g\gamma

\def\la{\left\langle}
\def\ra{\right\rangle}

\begin{document}

\title{Critical Casimir forces in a magnetic system: An experimental protocol}

\author{David Lopes Cardozo}
\author{Hugo Jacquin}
\author{Peter C. W. Holdsworth}
\affiliation{Laboratoire de Physique de l'\'Ecole Normale Sup\'erieure de Lyon, UMR CNRS 5672,
46 all\'ee d'Italie, 69007 Lyon, France}

\begin{abstract}

We numerically test an experimentally realizable method for the extraction of the critical Casimir force based on its thermodynamic definition as the derivative of the excess free energy with respect to system size. Free energy differences are estimated for different system sizes by integrating the order parameter along an isotherm. The method could be developed for experiments on magnetic systems and could give access to the critical Casimir force for any universality class. By choosing an applied field that opposes magnetic ordering at the boundaries, the Casimir force is found to increase by an order of magnitude over zero-field results. 
\end{abstract}

\maketitle

\section{Introduction}

Confinement of a critical system on the nanoscale leads to the critical Casimir force \cite{FG78}, whereby truncation of the diverging correlation length gives a singular contribution to the confining forces. This critical manifestation of the  Casimir force \cite{Ca48} has been of considerable interest 
over the last twenty years \cite{Krech94,G09}.
It has become accessible to measurement through a series of elegant experiments, probing either the forces on a localized colloidal particle \cite{HHGDB08}, or the Casimir contribution to a work function characterizing the thickness of a thin fluid film
\cite{GC02,GSGC06,FYP05,RBM07}. 
Theoretical \cite{ES94,K97,BU08,BAFG02} and numerical \cite{DK04,Hucht07,Hucht11,VGMD09,VD13,Hasenbusch09,Hasenbusch10a,Hasenbusch10b,Hasenbusch11,Hasenbusch13,Toldin13,toldin_critical_2010} studies however, systematically use approaches based on generalized thermodynamic relationships between the constraining forces and the relevant free energy. 
Here,  evolution of the free energy with system size yields the critical Casimir effect without direct access to the constraining force.

The search for experimental realizations of this effect has so far ignored magnetic systems - a surprising fact given that they have long been considered as the paradigm for studies of criticality (see, for example, Ref.\cite{S99}) and that the nano-engineering of magnetic thin films is particularly well-advanced \cite{VBL08}. In this paper, we numerically test a new protocol for measurement of the magnetic Casimir force based on the concept of generalized thermodynamic forces. This procedure could be adapted to experiments on magnetic thin films, 
or to systems  as diverse as ferroelectrics, liquid crystals or polymers and could give access to all universality classes including quantum criticality.

We concentrate on a system with scalar order parameter $m$, conjugate external field $h$,  volume $V$ and free energy $\Omega(T,h,V)$, close to a second order phase transition.
Anisotropic confinement is allowed for by setting $V=AL_z$, with $\sqrt{A}=L_{\parallel} \gg L_z$. We define dimensionless variables, $t=(T-T_c)/T_c$, $\tilde{h}=h/k_BT_c$, with $T_c$ the bulk three-dimensional critical temperature. 

For a magnetic system, $h$ is proportional to the applied magnetic field within an Ising description. In a simple fluid near the liquid gas critical point, $h\sim\mu-\mu_c$ is the chemical potential, measured with respect to the critical value, $\mu_c$ while near the de-mixing transition of a binary fluid, $h$ depends on the difference in chemical potential of the two species. Our analysis can easily be extended to include vector fields and order parameters, relevant for other universality classes such as, for example, helium films near the superfluid transition.

Strictly speaking, the thermodynamics of the magnetic system requires a fourth variable, $N$, the number of magnetic elements and hence a more general free energy, $\Omega(T,h,V,N)$. It becomes thermodynamically equivalent to the fluid systems by fixing the magnetic moment density $\rho={N\over{V}}$. In this case volume fluctuations impose fluctuations in the number of magnetic elements, so that one is dealing with a uniform magnetic medium.  While spontaneous fluctuations of this kind clearly cannot exist in conventional magnetic systems \cite{toldin_critical_2010}, the evolution of the free energy with system size can give indirect access to  the Casimir force and this is the subject of the present paper. 
An alternative constraint would be to impose $N$ constant, so that volume fluctuations would lead to magneto-elastic effects, as is the case in real magnets. In principle one could imagine magnetic experiments that directly measure Casimir forces through magneto-elastic coupling, although the separation of the critical and bulk contributions could be difficult. In practice, as magnetic exchange coupling varies rapidly with inter-atomic distance the critical properties are strongly perturbed and renormalization studies predict the transition to be driven first order by the coupling \cite{Bergman76}. This, in itself is an interesting field of study, but in the rest of the paper, we neglect all magneto-elastic effects and concentrate on the free energy which is generic to magnetic and fluid systems. For convenience, we set the microscopic length scale $\sigma=1$.

\section{Free energy and the critical Casmir force}

Neglecting surface corrections, the free energy near criticality takes the form $\Omega(T,h,V)=Vk_BT(\omega_a+\omega_s)$, where $\omega_{a/s}$ are the analytic and singular parts of the free energy density \cite{Goldenfeld92}. 
The critical Casimir effect is defined in the anisotropic confinement regime where the correlation length, $\xi$, lies in the range, $1 \ll \xi\sim L_z \ll L_{\parallel}$, so that  $\omega_s(t,\tilde{h},L_z^{-1})$. The $L_z$ dependence comes from the truncation of the correlation length near criticality \cite{Goldenfeld92} (dependence on the finite aspect ratio $L_z / L_{\parallel}$ \cite{Hucht11,toldin_critical_2010} is not considered in detail here).
The free energy can be developed to expose the contribution coming from this truncation:
\begin{equation}
\Omega(T,h,V)=  Vk_BT(\omega_a+\omega_s^0+\omega_s-\omega_s^0),
\end{equation}
where $\omega_a + \omega_s^0(t,\tilde{h})=\omega_{bulk}$ is the bulk free energy density in which the system is taken to the thermodynamic limit, $L_z\rightarrow \infty$, before the singular point, $t=0,h=0$ is approached, so that $\xi/L_z \rightarrow 0$ in all situations.
The difference, $Vk_BT (\omega_s - \omega_s^{ 0})$$=Vk_BT \omega_{ex}$ is referred to as the excess free energy between confined and bulk geometries \cite{Krech94,G09}.

In equilibrium and in the anisotropic limit defined above, the  confining  force per unit area is defined as
\begin{equation}
F_z=-{\frac 1A}\frac{\partial \Omega}{\partial L_z},
\label{force}
\end{equation}
so that the restriction of the critical fluctuations  introduces an anomalous term, 
 the critical Casimir force per unit area:
\begin{equation}
\begin{split}
f_c & =-k_B T \frac{\partial (L_z \omega_{ex})}{\partial L_z} \\
& =-k_BT\left(\omega_s-\omega_s^0+L_z {\frac{\partial \omega_s}{\partial L_z}}\right) \ .
\end{split}
\end{equation}

In fact, $L_z^{-1}$ plays an equivalent role in the criticality to reduced temperature and field, resulting in a third singular variable $Q={\frac{\partial V(\omega_s-\omega_s^0)}{\partial L_z^{-1}}}$, in analogy with the magnetic moment $M=Vm$ and the entropy $S$. The Casimir force, $f_c=k_B T L_z^{-1}(Q/V)$, is the natural physical observable related to this thermodynamics for which one finds a universal scaling form \cite{VD13}
\beq
f_{c} = k_BT L_z^{-d} \th \left( t L_z^{1/\nu},\tilde{h}L_z^{(\b+\g)/\nu} \right) \ .
\label{UnivScalForm}
\eeq
Here $d$ is the spatial dimension
 and critical exponents take their usual meaning \cite{Goldenfeld92}.

Extremely efficient numerical algorithms already exist for the simulation of the critical Casimir force within the framework of lattice based spin models. These algorithms make use of the thermodynamic relationship between force and free energy (\ref{force}), making a discrete estimate of $\displaystyle \frac{\partial \Omega}{\partial L_z}$, rather than simulating a direct force measurement. Free energy differences have been estimated by tracking the evolution of the excess internal energy with temperature for systems of size $L_z$ and $L_z-1$ \cite{Hucht07,Hucht11,Hasenbusch09,Hasenbusch10a,Hasenbusch10b}. Using this  method it has been possible to make accurate estimates of the scaling function extracted from work function measurements on helium films near the superfluid phase transition \cite{Hucht07}. It has also been successfully used to construct thermodynamic observables such as the singular contribution to the specific heat or order parameter \cite{Hasenbusch10a}. However, neither the internal energy at temperature $T$, nor  that at a required reference state \cite{Hasenbusch09} are themselves directly accessible in experiment.  In an alternate method  \cite{VGMD07}, direct access to free energy changes is achieved by adiabatically disconnecting a single layer of spins from a connected stack of $L_z$ layers: the coupling to the targeted layer and that between adjoining layers scale as $\lambda J$ and $(1-\lambda )J$ respectively for $0<\lambda< 1$. Integrating over $\lambda$, the internal energy difference between the coupled and decoupled system  allows an estimate of $\delta \Omega$ between systems of size $L_z$ and $L_z-1$. The explicit calculation of the free energy at a reference state can be avoided by subtracting results from two pairs of length scales. The method then provides  accurate estimates for the Casimir force for different universality classes and boundary conditions both for zero field \cite{VGMD09}, and more recently for non zero field \cite{VD13}.
Integration over the auxilary degree of freedom can be circumvented by equating the critical Casimir force with the anisotropic part of the  generalized internal stress tensor \cite{DK04}. The latter technique has been successfully used for varied situations, limited at present to zero field and periodic boundaries.

Given this success of spin models in the accurate computation of the critical Casimir force in almost all situations, it is paradoxical that no magnetic experiments exist which attempt to measure the scaling function from estimates of free energy differences.
The reason is that the above techniques, accurate though they may be, are not adapted to experiment. Here we show that equivalent results can be achieved by directly evaluating free energy changes through integration
from a reference state at high field into the critical region. This procedure is perfectly adapted to translation into the first experimental protocol for a magnetic system. 

The difference in free energy along an isotherm, between a reference state $(T,h_{0})$ and a final state $(T,h)$ is
\begin{equation}
\begin{split}
\Delta \Omega & = -\int^h_{h_{0}} M(T,h',L_z) dh' \ .
\end{split}
\end{equation}
Even if we choose $T\sim T_c$, if $|h_0|$ is chosen to be sufficiently large, the correlation length at the reference state will be small so that the reference free energy will be essentially that of the  bulk $\Omega(T,h_0,L_z)=Vk_B T \omega_{bulk}(T,h_0)$. As a consequence,  $\Delta \Omega$ should contain all the information of the Casimir effect at $(T,h)$.  A similar procedure could be developed along the temperature axis by integrating the entropy, $S(T)$, although the experimental observable is the specific heat, so that this route would require a double integration \cite{Hasenbusch10a}. Repeating this procedure for systems of size $L_z$ and $L_z-\delta L_z$ and applying the extensivity principle for the free energy away from criticality one finds
\begin{equation}
\begin{split}
\delta'\Omega (T,h,\ell) & \equiv \Delta\Omega(L_z) - \Delta\Omega(L_z-\delta L_z) \\
&= \delta \Omega(T,h,\ell) - \frac{\delta L_z}{L_z}\Omega(T,h_0,L_z)  \\
 &= \delta \Omega - \delta L_z A k_B T \omega_{bulk} ,
\end{split}
\end{equation}
where $\delta\Omega$ is the increment in free energy equating approximately  to $ \displaystyle \delta L_z \frac{\partial \Omega}{\partial L_z}$, evaluated at $h$ and $\ell=L_z-\delta L_z/2$. This intuitive choice has been shown rigorously to facilitate the approach to the scaling limit by minimizing the importance of corrections to scaling terms \cite{toldin_critical_2010}.
Non-critical surface free energy corrections cancel in the subtraction of the contributions from the two length scales. One now repeats the procedure for two sets of length scales  centered on $\ell$ and $\alpha \ell$. Subtracting results from the two pairs of length scales eliminates the free energy from the reference state, $\Omega(L_z,h^0)$, as well as the bulk contribution to the free energy at the point of interest, $\omega_s^0(t,h)$, providing a first estimate of the Casimir force:
\begin{eqnarray}
f_c^0(T,h,\ell)&=&-\left[\delta'\Omega (\ell)-\delta'\Omega (\alpha \ell)\right]\frac{1}{A\delta L_z}\nonumber \\
&=&-\left[\delta\Omega (\ell)-\delta\Omega (\alpha \ell)\right]\frac{1}{A\delta L_z} \nonumber \\
&\approx& f_c(\ell)-f_c(\alpha \ell) \ .
\end{eqnarray}
Given the universal scaling form for $f_c$  [Eq. (\ref{UnivScalForm})] one can define a scaling function for $f_c^0$
\begin{equation}
f_c^0(T,h,\ell)= k_BT\ell^{-d}\theta^0\left( u_t[\ell],u_h[\ell] \right) \ ,
\end{equation}
where $u_t=t \ell^{1/\nu}$ and $u_h=\tilde{h} \ell^{(\b+\g)/\nu}$ are the appropriate scaling variables. The scaling function $\theta^0(\ell)$ is related to $\theta$ at two different values of $u_t$ and $u_h$ by :
\begin{equation}
\theta^0(\ell) = \theta(\ell) - \alpha^{-d}\theta(\alpha \ell).
\label{theta0}
\end{equation}
Choosing $\alpha\approx 2$, the scaling function $\theta^0$ already provides a good estimate for the functional form of $\theta(u_t,u_h)$. To extract a complete estimate for $\theta(\ell)$, one can apply the procedure developed in Ref.  \cite{VGMD09} in which the approximate expression $\theta^n(\ell) = \theta^{n-1}(\ell) + \alpha^{-2^{n-1}d}\theta^{n-1}(\alpha^{2^{n-1}}  \ell)$ is iterated from $n=1$ to convergence (see Ref. \cite{toldin_critical_2010} and Appendix \ref{sec:Iteration}).

\begin{figure}[ht]
\includegraphics[width=8cm]{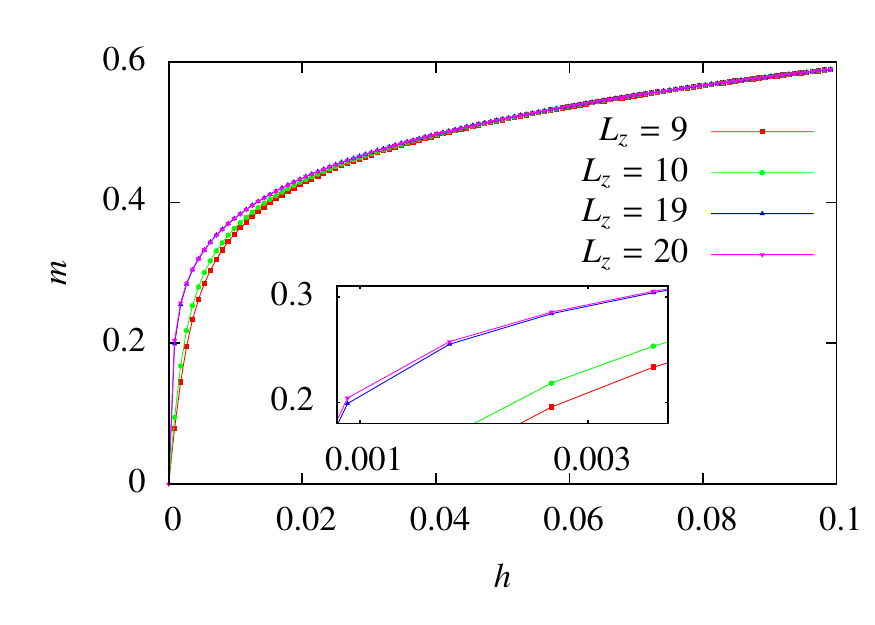}
\caption{(Color online) Magnetic order parameter vs $h$ at $T=T_c=4.5116J$ for $L_z=$9 (red squares), 10 (green dotts), 19 (blue triangles pointing up), 20 (magenta triangles pointing down) and $A=3600$ for periodic boundary conditions. \textbf{(Inset)} Blow-up of the low-field region of the magnetization. 
}
\label{magnetization_PBC}
\end{figure}

\section{Results}

We have tested these ideas through Monte Carlo simulation of a nearest neighbor Ising spin system with coupling strength $J$ and external field $h$, on a cubic lattice with $L_{\parallel}>L_z$, periodic boundaries in the $\hat{x}-\hat{y}$ plane and varying boundaries along the $\hat{z}$ axis. The Hamiltonian reads
\beq
H = - J \sum_{\la i , j \ra} s_i s_j - h \sum_i s_i \ ,
\eeq
where $\la i,j \ra$ denotes a sum over nearest neighbors, $s_i=\pm 1$ and the sum runs from $i=1,N$ ($N=V=L_zA$). The magnetic order parameter is then
\beq
m=\frac{1}{V}\left\langle \sum_i s_i\right\rangle,
\eeq
where $\left\langle X \right\rangle$ is a thermal average. We have used the Wolff algorithm, adapted to work in the presence of a symmetry breaking field \cite{DRORT92} (see Appendix \ref{sec:MCstepAndError}). For simplicity, $J=1$ in our simulations.

In Fig. \ref{magnetization_PBC} we show the evolution of the magnetization with applied field for $T=T_c$ for $L_{z}=9, 10, 19, 20$ and for periodic boundaries along $\hat{z}$ (Appendix \ref{sec:SystemSizes} give comments on the choice of system sizes). Similar results are obtained for ($+,+$) and $(+,-)$ boundaries, where spins on the boundaries are fixed in the same, or in opposite directions. The difference  in $m(L_z,h)$, for small $h$ is clearly visible for $L_z=9$ and $10$ becoming much smaller for the larger $L_z$. The Casimir force comes from the integral of these differences with field, so that system sizes straddling $L_z=10$ appear to offer a good pragmatic place to start.
For this length scale the effect is pronounced, while one is already in the scaling regime to within a reasonable approximation. In addition, magnetic films of this thickness can be produced with great precision so that these parameters already correspond to the state of the art for thin film production \cite{VBL08}. 

\begin{figure}[ht]
\centering
\includegraphics[width=8cm]{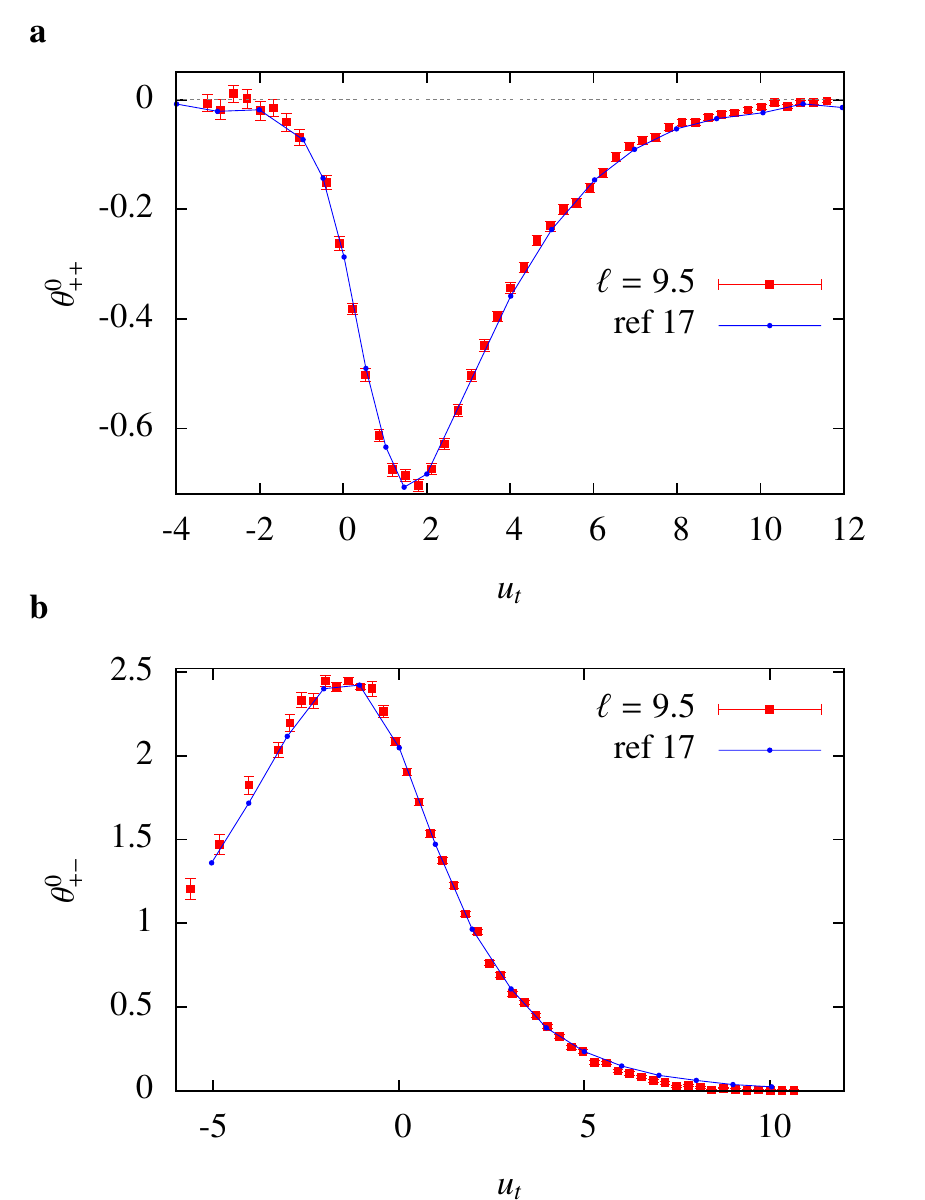}
\caption{(Color online) Zeroth order scaling function $\theta^0$ vs $u_t=t\ell^{1/\nu}$ for $h=0$. Data from the magnetic protocol outlined in the text (red cross), data from Ref. \cite{VGMD09} (blue line). Figure {\bf a)}  $(+,+)$ boundaries, {\bf b)}  $(+,-)$. In all cases $\ell=9.5$, $\delta L_z=1$ and $\a \ell = 19.5$, while $A=3600$. The error bars were computed using a modified bootstrap method and an estimate of the autocorrelation time (see Appendix \ref{sec:MCstepAndError}).}
\label{theta0_plusplus}
\end{figure}

In Fig. \ref{theta0_plusplus} we compare the zeroth order scaling function, $\theta^0(u_t,0)$ extracted using the magnetic protocol described above with that from reference \cite{VGMD09} for (a) $(+,+)$  and (b) $(+,-)$ boundary conditions. 
In all cases $\ell =9.5$, $\delta L_z=1$ and $\alpha \ell=19.5$. At each temperature the value of $h_0$ characterizing the reference state was chosen large enough so that $\theta^0$ approached an asymptote (see Appendix \ref{sec:h0andIntegration}). One can observe excellent agreement between the two data sets for both boundary conditions, thus confirming our protocol as a viable method of extracting critical Casimir forces. We have also successfully tested our protocol against the adiabatic method for periodic boundaries.
The difference in sign and amplitude of the Casimir force  between $(+,+)$ and $(+,-)$ boundaries has its origin in the excess entropy of the trapped interface. This spectacular inversion and scale change is perfectly captured by our thermodynamic protocol. From here, the universal function $\theta$ can be extracted by iteratively solving Eq.(\ref{theta0}).

Arriving at a scale free function from these system sizes also requires a delicate analysis of corrections to scaling \cite{Ha12}. Having made contact with previous work for these modest system sizes, we account for the corrections here by rescaling the data to the universal scaling amplitude, $\theta(0,0)=2\Delta$ for each set of boundaries. 
If this technique were developed in magnetic thin film experiments, it is likely that initial measurements would require scaling in the same way, as was the case for early experimental data for helium films to remove amplitude shifts due to uncontrolled surface roughness \cite{GSGC06,Garcia99}. Numerical estimates given in the literature vary: $\Delta_{++}=-0.376(29)$ and $\Delta_{+-}=2.71(2)$ \cite{VGMD09}, $\Delta_{++}=-0.410(7)$ and $\Delta_{+-}=2.806(10)$ \cite{Hasenbusch10b}. Here we take values from \cite{VGMD09}, as our method relates to this work. We return to this subject below, where we present some initial finite size scaling results for the critical Casimir force in finite field.

\begin{figure}[ht]
\begin{center}
\includegraphics[width=8cm]{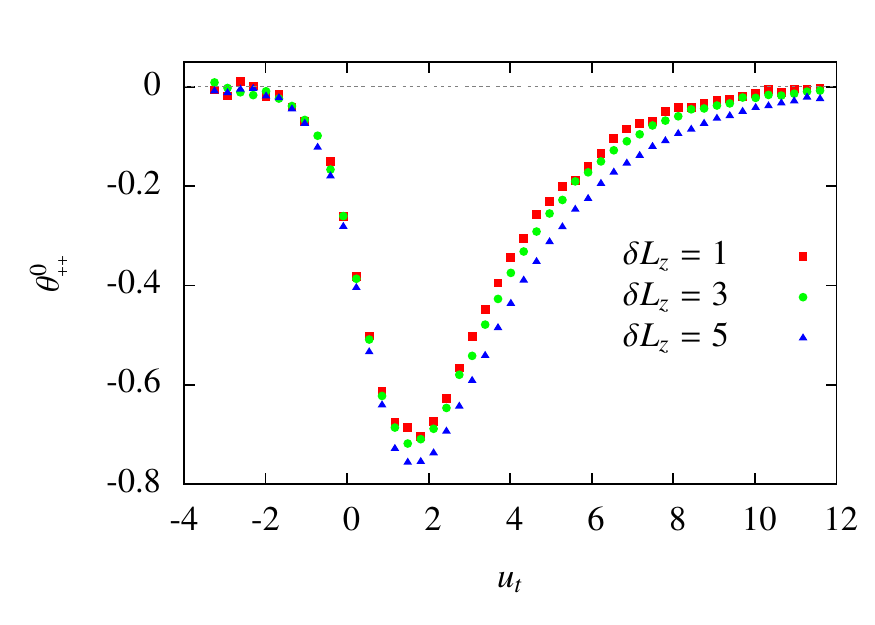}
\caption{(Color online) Scaling function $\theta^0$ vs $u_t=t\ell^{1/\nu}$ for $h=0$, $\ell=9.5$ and $(+,+)$ boundaries. Data from the magnetic protocol outlined in the text (red squares) for $\delta L_z =1$, $L_z=10$, for $\delta L_z =3$, $L_z=11$ (green dots), for $\delta L_z =5$, $L_z=12$ (blue triangles), with $A=3600$.}
\label{theta-DL} 
\end{center}
\end{figure}

The experimental feasibility of this protocol requires the fabrication of samples with thickness resolution better than $\delta L_z$ as well as the capacity to keep the uncontrolled errors generated by measurements on different samples at different times below the same threshold. The chances of success would clearly be increased if one could increase $\delta L_z$ above a monolayer. With this in mind we have investigated the measured Casimir effect for different values of $\delta L_z$. 
The results are shown in Fig. \ref{theta-DL} for $\delta L_z=1, 3$ and $5$, for fixed $\ell=9.5$. Remarkably, the evolution of the estimated function, $\theta^0$, on moving from $\delta L_z=1$ to $3$ is extremely small, with a typical difference of less than $5\%$ as the function passes through its minimum between $u_t=1$ and $u_t=2$. This small evolution is only just resolvable above the statistical error on our data, which is approximately $1.5\%$ in this region.  Even for $\delta L_z=5$ the evolution remains less than $11\%$ around the minimum of the function, while in all cases, increasing $\delta L_z$ enhances the measured Casimir force. In addition, as the free energy difference $\delta \Omega$ increases with $\delta L_z$, the statistical errors are reduced, even in the wings of the figure. The effect therefore appears extremely robust and our results strongly suggest that it would stand up to the technical problems  encountered in dedicated experiments on magnetic thin films.

\begin{figure}[ht]
\begin{center}
\includegraphics[width=8cm]{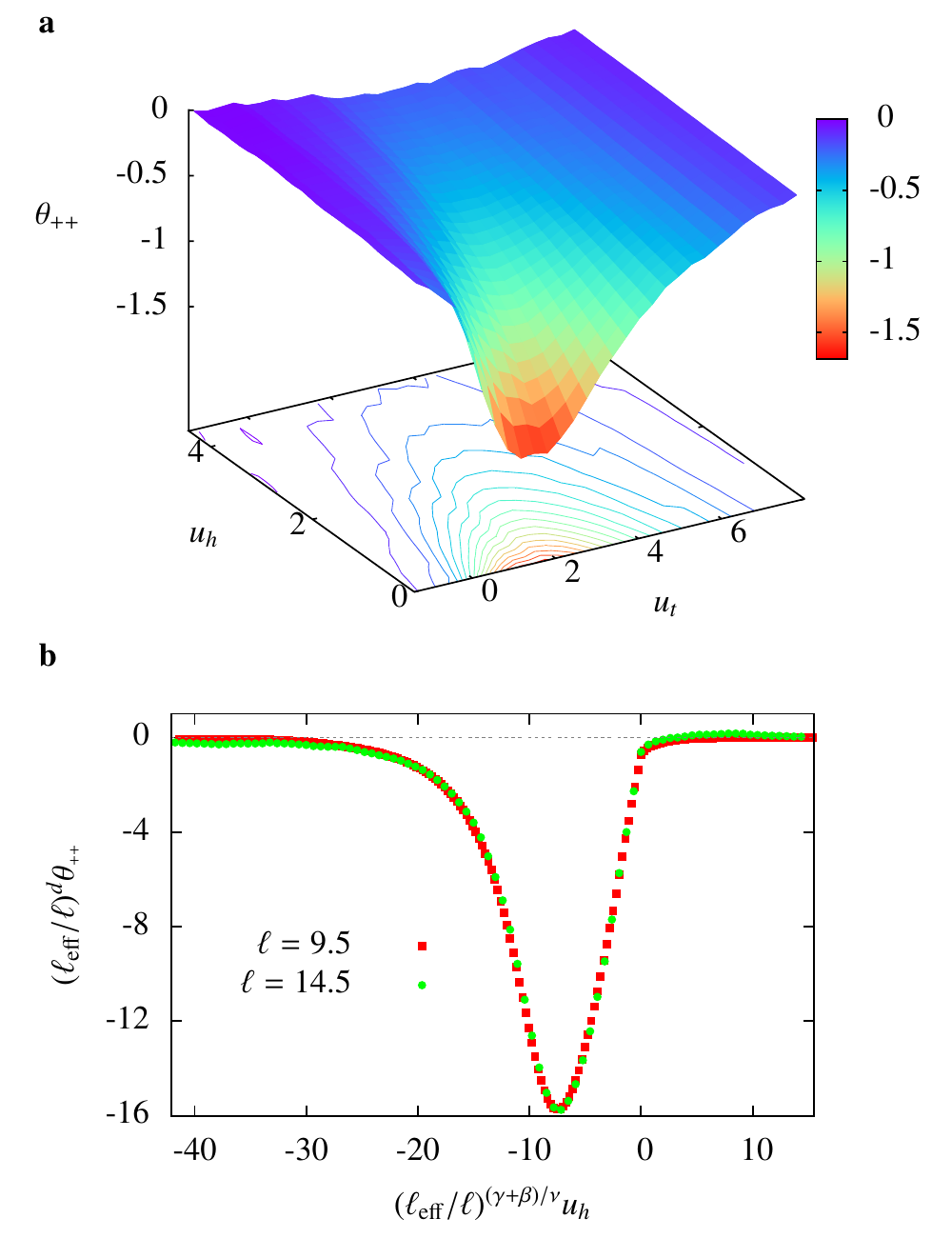}
\caption{(Color online) {\bf a)} $\theta(u_t,u_h)$ for $(+,+)$ boundaries, found using the magnetic protocol outlined in the text for $\ell=9.5$, $\alpha\ell=19.5$, $\delta L_z=1$ and $A=3600$. The field is confined to the $+$ direction. The function was scaled to universal amplitude, $\theta(0,0)=2\Delta_{++}=-0.75$. The lines projected onto the base show contours of equal Casimir force. \newline {\bf b)} $(\ell_{\rm eff}/\ell)^d\theta_{\scriptscriptstyle ++}(0,(\ell_{\rm eff}/\ell)^{(\gamma+\beta)/\nu}u_h)$ for $(+,+)$ boundaries under the same conditions, with field spanning both $+$ and $-$ directions. Two sets of system sizes were used: $\ell=9.5$, $\alpha \ell=19.5$ (red squares) and $\ell=14.5$,$\alpha \ell=29.5$ (green dots). The data sets were rescaled to universal amplitude and width by replacing $\ell$ with $\ell_{eff}=\ell+\delta \ell$, with $\delta \ell = 2.8$, as detailed in Appendix \ref{sec:leff}.}
\label{theta-t-h} 
\end{center}
\end{figure}

Until recently \cite{VD13,Vasilyev14} there has been only minimal interest in the scaling of the critical Casimir force along the field axis. This can be explained in part by an absence of experimental motivation as it is difficult to probe the field variable in present setups: for the superfluid transition in ${}^4$He films \cite{GC02,GSGC06}, $h$ is not accessible, while for binary liquid films \cite{FYP05}, experiments are performed for fixed concentrations, rather than conjugate field.
However, experiments on thin film magnets lend themselves naturally to critical scaling in both $u_t$ and $u_h$. Our numerical protocol is equally well adapted and is in fact, particularly efficient, as all points along an isotherm contribute to $\theta(u_t,u_h)$. Our procedure therefore opens up a new direction for the study of these forces. In Fig. \ref{theta-t-h}$a$ we show $\theta(u_t,u_h)$ for $(+,+)$ boundaries, illustrating the form of the scaling function in the half plane, $h>0$. This figure requires the same computational effort as the one dimensional data sets shown in Fig. \ref{theta0_plusplus}. 

The scaling function shows no minimum value as a function of field. The minimum can be found in the half plane, $h<0$, with the field in the opposite direction to the pinned boundaries. Remarkably, as we show in Fig. \ref{theta-t-h}$b$, $\theta$ plunges to values more than an order of magnitude lower, as one crosses the line to negative field values. This unexpectedly large amplitude \cite{VD13} comes from the competition between opposing surface and bulk fields. At large separation, the applied field  imposes two magnetization interfaces. For smaller $L_z$, this frustration is lifted and symmetry is broken in the direction of the boundary field resulting in a particularly large Casimir force, which could be accessed in thin film experiments. 

We show in Fig. \ref{theta-t-h}$b$ data for two values of $\ell$. The collapsed data are the result of a procedure allowing the anticipation of corrections to scaling and an extrapolation of the measured function $\theta(\ell)$ to the scaling function of the thermodynamic limit \cite{toldin_critical_2010,Ha12,VD13}. In this scheme an effective length, $\ell_{eff}=\ell+\delta \ell$, replaces $\ell$, with $\delta \ell$ chosen to collapse the data. The parameter $\alpha$ used in the iteration procedure varies in consequence; $\alpha_{\rm eff}=\frac{\alpha \ell + \delta \ell}{\ell + \delta \ell}$. 
The process has been shown to capture corrections to scaling in a controlled manner in the Blume-Capel model \cite{toldin_critical_2010}, but is used here in an exploratory manner. 
A single value, $\delta \ell=2.8$ leads to good data collapse and a reasonable estimate for the universal scaling function. From the scaled data we find $\Delta_{\scriptscriptstyle ++}(\ell=9.5)=(\ell_{\rm eff}/\ell)^d\theta_{\scriptscriptstyle ++}(0,0)/2=-0.30(3)$ and $\Delta_{\scriptscriptstyle ++}(\ell=14.5)=(\ell_{\rm eff}/\ell)^d\theta_{\scriptscriptstyle ++}(0,0)/2=-0.36(7)$. Given that these estimates are taken from scaling curves of considerably larger amplitude that those in the half plane for positive field, they appear in acceptable agreement with previously found values  \cite{VGMD09,Hasenbusch10b}. The rescaling process  and the iteration process specific to this case are discussed in more detail in Appendixes \ref{sec:Iteration} and \ref{sec:leff}.

\section{Discussion} Having established the potential of the method to construct the Casimir scaling function from measurements of the magnetic moment, we now return to confrontation with experiment. Perhaps the most important point to address is the scale of the magnetic field required. Most of the Casimir signal comes from small fields, but in order to evacuate the entire Casimir effect it was necessary to go to fields as large as $|h_0|/J\sim 0.3$ (see Fig. \ref{magnetization_PBC}).  
One is therefore limited to ferromagnets with Curie temperature up to around $30$ K. 
Experimental systems \cite{DJM73} potentially cover a wide range of universality classes and surface conditions, opening the possibility for a rich variation in universal behavior. Our protocol can easily be extended to cover many of these situations.  
Other universality classes  can easily be treated, as can the anisotropic spin Hamiltonians often appearing in magnetic systems. In such cases one expects crossover from the microscopic starting point to the final universality class as the correlation length grows. These effects could be studied in detail and could be highly relevant for magnetic experiments. Boundary effects could be extended to include both rough and soft interfaces \cite{GSGC06,Garcia99}. However, 
materials with a strongly anisotropic spin Hamiltonian and hard smooth interfaces offer the most promising starting point.

Magnetic materials show essentially perfect model magnetism in many instances (see, for example, Refs. \cite{IH74,RMH99,ABH93}).
Candidates for the Casimir effect would be ferromagnets and could include both metallic and insulating materials. Promising characteristics  that one might consider include: in iron doped palladium films both the transition temperature  and film thickness can be accurately controlled \cite{KAH12}, while insulating compounds Tb(OH)$_3$ and K$_2$CuCl$_4$:2h$_2$0 are examples of Ising and Heisenberg ferromagnets respectively with Curie temperatures in the $5$-K range \cite{DJM73}. The metallic RKKY material, HoRh$_4$B$_4$ is a perfect mean-field ferromagnet \cite{OKO82} which could offer access to mean-field critical Casimir forces for the first time. The dipolar ferromagnet, LiHoF$_4$ is the archetypical transverse field Ising system \cite{BRA96} which, if produced as a film could provide a candidate for the study of Casimir forces at a quantum critical point \cite{Sachdev11}. Finally, we remark that our protocol could be extended to study non magnetic systems such as ferroelectrics, liquid crystals or simple and binary fluids, as it offers a generic method when the field conjugate to the order parameter is a control parameter. It could then be experimentally relevant in setups for fluid systems if  the chemical potentials could be controlled, rather than the concentrations.

\begin{acknowledgments}
We thank S. Ciliberto for comments and support throughout this project, D. Bartolo, S. T. Bramwell,  C. Charles, M.J.P. Gingras, B. Hjorvarsson, F. Parisen Toldin and H. Zabel for useful discussions and A. Gambassi for authorizing use of data from Ref. \cite{VGMD09}. The work was financed by the ERC grant OUTEFLUCOP and used the numerical resources of the PSMN at the ENS Lyon. P.C.W.H. acknowledges financial support from the Institut Universitaire de France.  
\end{acknowledgments}

%%%%%%%%%%%%%%%%%%%%%%%%%%%%%%%%%%%%%%%%%%%%%%%%%%%%%%%%%%%%%%%%%%%%%%%%%%%%%%%%%%%%%%%%%%%%%
\appendix

\section{The Monte-Carlo step and error analysis}
\label{sec:MCstepAndError}

In this appendix, we briefly describe the Monte Carlo algorithm we used and the definition of the Monte Carlo step. The precision of the simulation and error analysis are also discussed.

We have used the Wolff algorithm  \cite{newman_barkema} to simulate Ising systems to  reduce critical slowing  down in the critical region. A Monte Carlo step was defined by first computing  the mean size of clusters generated by the Wolff algorithm $ \left\langle C \right\rangle$ at each temperature and $h=0$. One Monte Carlo step is then composed of $\frac{AL_z}{\left\langle C \right\rangle}$ calls to the Wolff algorithm, so that, on average, $AL_z$ spin flips are performed during each step. To include a magnetic field in the simulation, spin clusters are created in the same way as for the Wolff algorithm at zero magnetic field but the clusters are no longer systematically flipped. 
We chose to use the ''ghost spin'' method\cite{LR89} \cite{Wa89} in which each spin of a cluster can be linked to a ghost spin of fixed value $\sigma_{ghost}=+1$ representing the magnetic field $h$. The probability of coupling a spin $\sigma$ belonging to the cluster to the ghost spin is $1-e^{-2\beta \sigma h}$ if $\sigma h > 0$ and $0$ otherwise: any cluster linked at least once to the ghost spin is left unflipped. Each time a spin is added to a cluster it is possible to test whether this spin couples to the ghost spin or not. In the case that it does the growth of the cluster is stopped to save computational time. We simulated the Ising model on a cubic lattice with either complete periodic boundary conditions, or periodic boundaries in the $\hat{x}$ and $\hat{y}$ directions and closed $(+,+)$ and $(+,-)$ boundary conditions along the $\hat{z}$ direction. The fixed boundary conditions can be considered as local magnetic fields. For temperatures below $T_c$, as the absolute value of the magnetic field $|h|$ increases, the number of rejected cluster flips increases dramatically, resulting in an increase of the autocorrelation time and therefore a loss of efficiency of the algorithm. Obtaining precise results at low temperature, particularly for $(+,-)$ boundary conditions \cite{VGMD09}, requires a particularly large computation time.
The data we present in the article were obtained using a number of Monte Carlo steps ranging from $5 \times 10^4$ for $(+,+)$ boundary conditions at the higher temperatures to $7 \times 10^7$ for temperatures far below $T_c$ in systems with $(+,-)$ boundary conditions where the efficiency of the algorithm is at its lowest. 

The statistical error is evaluated using a modified bootstrap method \cite{newman_barkema}. As the presence of fixed boundary conditions and bulk magnetic field increases  the correlation time $\tau_{\rm corr}$ dramatically, we interpret the bootstrap method as providing a value for 
$\sigma_{m}/\sqrt{N_{\rm step}}$ where $\sigma_{m}^2=\left\langle m^2 \right\rangle-\left\langle m \right\rangle^2$ is the variance and $N_{\rm step}$ is the number of Monte Carlo steps performed, rather than the error itself. To compute the statistical error we estimated the autocorrelation time $\tau_{\rm corr}$ and then take the error to be $\sqrt{2\tau_{\rm corr}\sigma_{m}^2/N_{\rm step}}$ \:\cite{newman_barkema}.

\section{Choice of the system sizes}
\label{sec:SystemSizes}

In this appendix, we summarize some of the important aspects that have to be taken into account when choosing system sizes.

We chose to study preferentially system thicknesses $\ell=9.5$ and $\a \ell=19.5$ and initially take $\delta L_z=1$. Different constraints motivate this choice: first $\ell$ has to be big enough with respect to the variation $\delta L_z$ so that the derivative of the free energy with respect to the system size can be safely approximated by the differential $\frac{\delta \Omega}{\delta L_z}$ (as discussed in the main text, $\delta L_z=3$ and $5$ have also been studied in order to test the robustness of the approach). Secondly, $\ell$ must be big enough to allow an approach into the three dimensional scaling regime.
This choice is moderated by the fact that the difference in magnetization for different system sizes falls to zero as the scaling limit is approached, so that a pragmatic compromise is required, both in simulation and in any future experiment. These considerations motivated our choice of the relatively modest system size, $\ell=9.5$ for many of the results presented.
Thirdly, $\a$ has to be as large as possible to have a fast convergence of the iteration process that extracts the approximation $\theta^k$ from the measured $\theta^0$. Here $\a = 19.5/9.5 \approx 2$. 
$\sqrt{A}=L_{\parallel}$ should be chosen as big as possible with respect to $L_z$ in order to ensure that we stay in the anisotropic confinement regime. In all our simulations we used $A=3600$ enabling us to directly compare our results with those from Ref. \citenum{VGMD09} where one can find detailled discussions on the impact of system size and of  corrections to scaling, on the form of the universal function of the Casimir force obtained in the Ising and XY models. \\

\section{Choice of $h_{0}$ and integration procedure}
\label{sec:h0andIntegration}

To be able to extract the free-energy by integration of the order parameter, it is necessary to chose a suitable reference magnetic field $h_0$. We define here the function $D(T,h,\ell)$ that enables us to make such a choice.

Figure \ref{magnetization_PBC} shows the magnetic order parameter as a function of magnetic field for four different systems sizes $L_z$ with periodic boundary conditions at $u_t=0$. At low magnetic field the four curves do not superimpose showing clearly the finite-size effect that we want to capture. At zero magnetic field the value of the magnetization $m(h=0)=0$ is imposed by magnetic field reversal  symmetry. At low magnetic field the magnetization depends on $L_z$ but as the magnetic field is increased the curves asymptotically merge. 

\begin{figure}
\centering
\includegraphics[width=8cm]{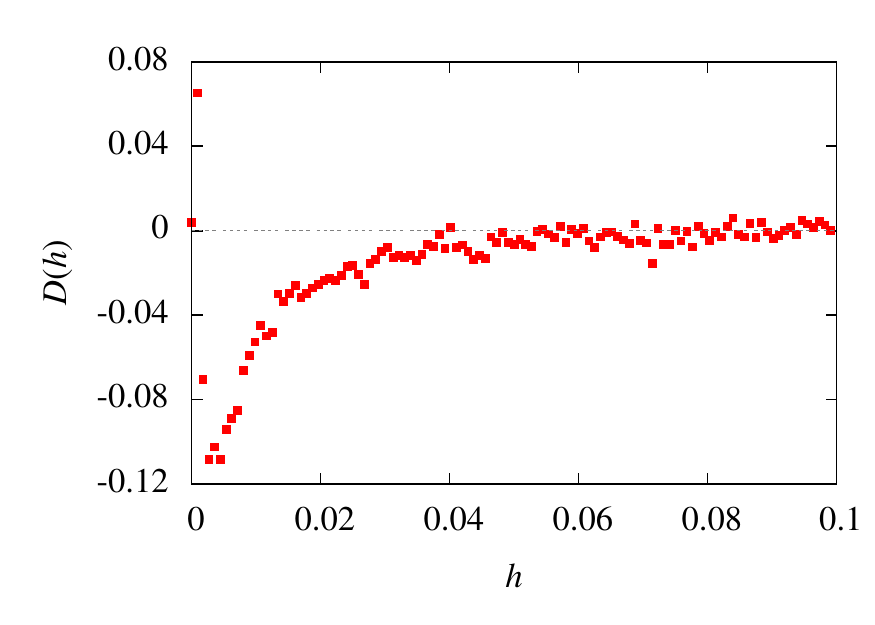}
\caption{(Color online) Function $D(T,h,\ell)$ defined in Eq.(\ref{def_D}) with respect to the magnetic field h. The data were obtained at $T=T_c$ for periodic boundary conditions and $\ell=9.5$,$\a \ell=19.5$ and $A=3600$ (same data as in Fig.\ref{magnetization_PBC}). The integration of $D(T,h,\ell)$ over $h$ gives $\theta^{0}$. $D(T,h,\ell)$ goes to zero as $h$ is increased, $h_{0}$ should be chosen so that $D(T,h,\ell)$ is zero within the current precision of the simulation, ensuring that finite size effects are suppressed by this field.}
\label{DeltaDelta_PBC}
\end{figure}

Let us define the function :
\begin{widetext}
\beq 
D(T,h,\ell) = \frac{1}{A\delta L_z} \left[ M\left(T,h,\a \ell+\frac{\delta L_z}{2}\right)- M\left(T,h,\a \ell-\frac{\delta L_z}{2}\right)- M\left(T,h,\ell+\frac{\delta L_z}{2}\right)+ M\left(T,h,\ell-\frac{\delta L_z}{2}\right) \right]  \ ,
\label{def_D}
\eeq
\end{widetext}
so that : 
\beq
\theta^{0}(u_t,u_h) = L_z^d \b \int_h^{h_{0}} dh' ~ D(T,h',\ell) \ .
\label{integ_D}
\eeq
Functions $D$ and $\theta^{0}$ also depend on the choice of the parameters $\alpha$ and $\delta L_z$  but we omit this dependencies in our notations for sake of lightness.
$D(T,h,\ell)$ can be used to find a suitable reference magnetic field $h_{0}$ such that $D(T,h_{0},\ell) \approx 0$. Fig.\ref{DeltaDelta_PBC} shows $D(T,h,\ell)$ computed with the data presented in Fig.\ref{magnetization_PBC}.  We see that it goes to zero as $h$ is increased, enabling us to chose a suitable reference magnetic field $h_{0}$ which suppresses completely the finite size effect within the current precision of the simulation. As the size of the critical region in the $h$ direction changes with the temperature, the reference magnetic field also varies and $h_{0} \underset{|t| \to +\infty}{\to} 0$.
After choosing a suitable $h_0$ the integration of Eq.(\ref{integ_D}) was performed using Simpson's rule from $h_{0}$ to $h$ for all computed values of $h$.

\section{Iteration procedure}
\label{sec:Iteration}

In Eq. (\ref{theta0}) we show the relation between the zeroth order scaling function $\theta^0$ and the scaling function of the Casimir force $\theta$ itself. 
Extending the method of Ref. \citenum{VGMD09} to the case of the Casimir force with a magnetic field, Eq.(\ref{theta0}) can be solved iteratively to extract the function $\theta(u_t,u_h)$ from the measured quantity $\theta^{0}(u_t,u_h)$. If $\a$ is chosen greater than 1, as was the case in our simulations, we can consider, as a first approximation to the function $\theta(u_t,u_h)$: 
\beq
\theta^0 (u_t,u_h) \approx \theta(u_t,u_h) \ .
\eeq 
Let us now consider the following recursion relation to higher-order approximations of $\theta(u_t,u_h)$ :
\beq \begin{split}
\theta^{n \ge 1} (u_t,u_h) &= \theta^{n-1}(u_t,u_h) \\ +  & \a^{-2^{n-1}d}\theta^{n-1} (\a^{2^{n-1}/\nu}u_t,\a^{2^{n-1}(\beta+\gamma)/\nu}u_h)  \ .
\label{recursion}
\end{split}
\eeq 
Rewritting this relation as a recursion procedure for the function $\theta^{0}(u_t,u_h)$ only, we can show that it converges toward :
\beq \begin{split}
\hat{\theta} (u_t,u_h)&=\underset{n\to \infty }{\lim} \hat{\theta}^{n} (u_t,u_h) \\
& = \sum_{n=0}^{\infty} \a^{-dn} \theta^{0}(\a^{n/\nu}u_t,\a^{n(\beta+\gamma)/\nu}u_h) \ .
\end{split} \eeq 
The series defining $\hat{\theta}(u_t,u_h)$ converges because $\a^{-dn}$ decays exponentially with $n$ and $\theta^{0}(u_t,u_h)$ is expected to be bounded, having a finite maximum close to the critical point and decaying exponentially quickly for $u_t,u_h \to \pm \infty$.
By injecting the expression of function $\hat{\theta}(u_t,u_h)$ into Eq.(\ref{theta0}) we see that it is indeed a solution to the equation. A finite number of iterations therefore provides an approximation $\theta^{n}(u_t,u_h)$ for the universal scaling function $\theta(u_t,u_h)$. 
\begin{figure}
\centering
\includegraphics[width=8cm]{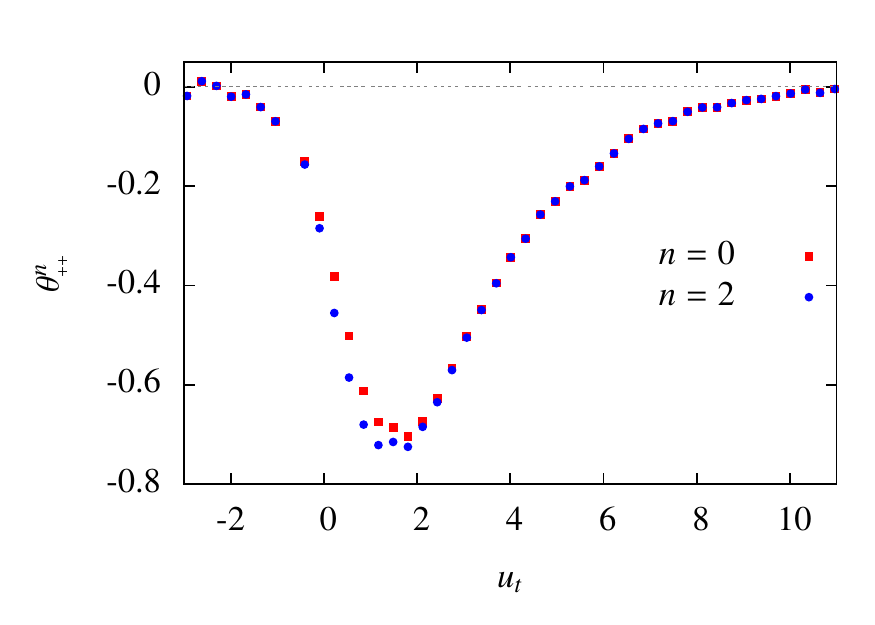}
\caption{(Color online) Evolution of the scaling function with the iteration procedure. Function $\theta^{n}_{++}(u_t,0)$ of the Casimir force for $n=0$ (red squares) and $n=2$ (blue dots) as a function of reduced variable $u_t=t ~ L_z^{1/\nu}$, computed using the proposed integration method for (++) boundaries and $\ell=9.5$, $\alpha\ell=19.5$,$\delta L_z=1$ and $A=3600$.}
\label{g_theta}
\end{figure}

This iterative process converges rather quickly: for a typical value of $\alpha=2$ in three dimensions, for $n=3$ we already have : $\a^{-2^{n-1}d} \sim 10^{-4}$, 
$\a^{2^{n-1}/\nu} \sim 10^{2} $,
$\a^{2^{n-1}(\beta+\gamma)/\nu} \sim 10^{3}$. The correction given by the fifth iteration is therefore expected to be small given the very small value of the parameter $\a^{2^{n-1}}$ and that the point $(\a^{2^{n-1}/\nu}u_t,\a^{2^{n-1}(\beta+\gamma)/\nu}u_h)$ reached will be far from the critical point, except for extremely small values of $(u_t,u_h)$.
Note that using this recursion relation to obtain $\theta$ over a given range of $u_t$ and $u_h$ values requires that the function $\theta^{0}$ is measured over a much wider range, since each iteration dilutes the chosen window. Further, the procedure requires the use of
 values for  $\theta^{0}(u_t,u_h)$  over the continuous range of variables, not just the discrete set used in the Monte-Calo simulation. These values are estimated using spline interpolation of the computed values of $\theta^{0}$. In practice we have chosen $\a \approx 2$, and have used two iterations to obtain an estimate of $\theta$. For $n=3$, we found  that all points $(\a^{2^{n-1}/\nu}u_t,\a^{2^{n-1}(\beta+\gamma)/\nu}u_h)$ (except for $(u_t=0,u_h=0)$ of course) fall outside the range of values of $(u_t,u_h)$ used in our Monte Carlo simulation. Hence, their contribution could safely be considered to be negligible within the precision of our simulation. Figure \ref{g_theta} presents the evolution of $\theta^{n}_{++}(u_t,0)$ between $n=0$ and $n=2$, with data found using $\ell=9.5$, $\alpha\ell=19.5$, $\delta L_z=1$ and $A=3600$.

\begin{figure}
\includegraphics[width=8cm]{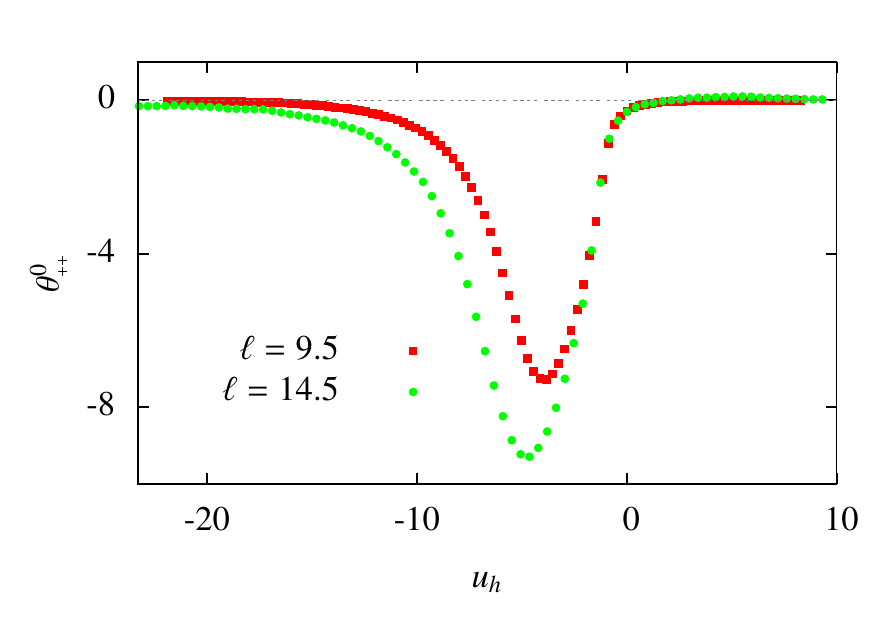}
\caption{(Color online) Zeroth order scaling function $\theta^0_{\scriptscriptstyle ++}(0,u_h)$ for (++) boundaries obtained with two different sets of system sizes: $\ell=9.5$, $\alpha \ell=19.5$ (red squares) and $\ell=14.5$, $\alpha \ell = 29.5$ (green dots). For both sets of data $\delta L_z=1$, $A=3600$ and $\alpha \approx 2$.}
\label{before_Leff}
\end{figure}

In the case of finite field and $+,+$ boundaries we encounter a large amplitude Casimir force for a field in the reverse direction, $h<0$, as discussed in the main text and shown in Fig. \ref{before_Leff}. This puts a strain on the iteration procedure in the region where the scaling function evolves most rapidly with field, producing a kink in the estimated function $\theta(0,u_h)$ for small, negative $h$. Results are  shown in Fig. \ref{theta-t-h-alone} for two system sizes, $\ell=9.5$ and $\ell=14.5$. The kink appears less pronounced for the larger system size, which suggests that it is an artifact of the procedure for small systems. More work is required to confirm this point.

\section{Rescaling of $\theta$ : choice of $\ell_{\rm eff}$}
\label{sec:leff}

\begin{figure}
\begin{center}
\includegraphics[width=8cm]{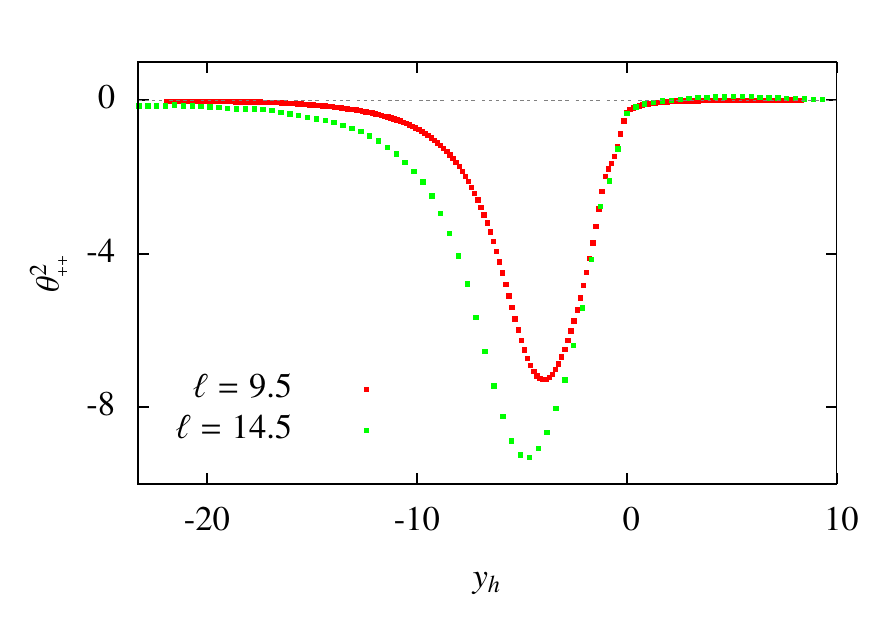}
\caption{(Color online) Function $\theta^2_{\scriptscriptstyle ++}(0,u_h)$ for $(+,+)$ boundary conditions and different system sizes, with field spanning both $+$ and$-$ directions. The function was obtained after applying the iteration procedure described in the text twice, that is to say to convergence within our current precision. The corresponding functions $\theta^0$ are displayed in Fig. \ref{before_Leff} . Data were obtained with two sets of systems sizes $\ell=9.5$, $\alpha\ell=19.5$ and $\ell=14.5$, $\alpha\ell=29.5$ with $(+,+)$ boundaries, $\delta L_z=1$ and $A=3600$.}
\label{theta-t-h-alone} 
\end{center}
\end{figure}

\begin{figure}
\includegraphics[width=8cm]{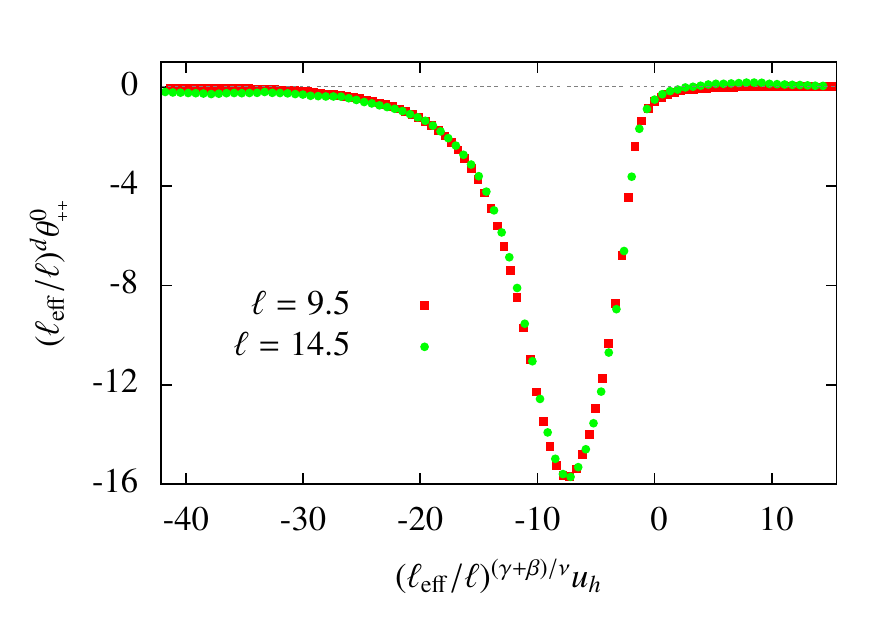}
\caption{(Color online) Zeroth order scaling function with $(+,+)$ boundaries for $\ell=9.5$, $\alpha \ell=19.5$ (red squares) and $\ell=14.5$, $\alpha \ell = 29.5$ (green dots) collapsed using an effective length scale $\ell_{\rm eff}=\ell+\delta \ell$. Here  $\delta \ell = 2.8$. This correction affects both the amplitude of the function by a factor of $(\ell_{\rm eff}/\ell)^d$ and the reduced parameter $u_h$ by a factor $(\ell_{\rm eff}/\ell)^{(\gamma+\beta)/\nu}$. For all data $\delta L_z=1$ and $A=3600$.}
\label{choice_Leff}
\end{figure}
\begin{figure}
\includegraphics[width=8cm]{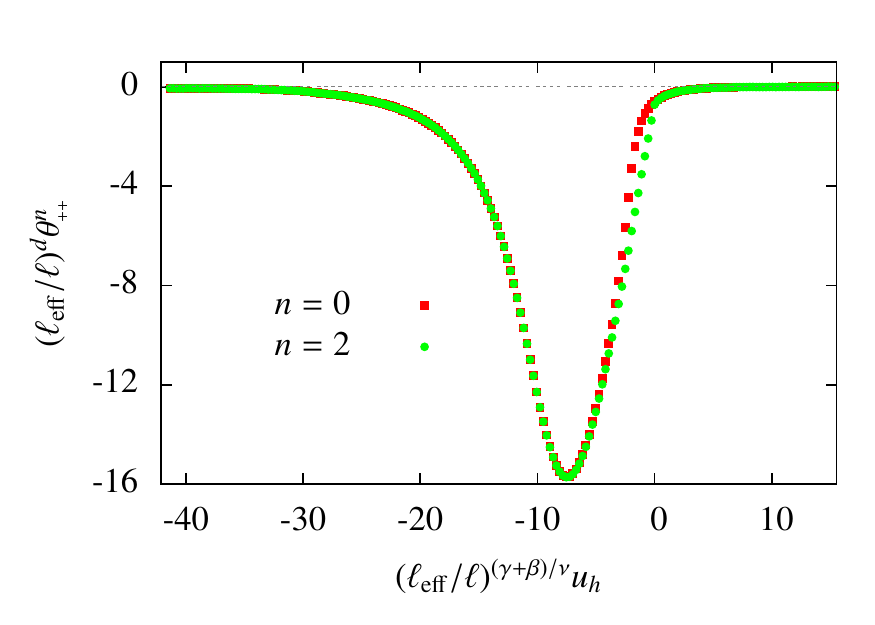}
\caption{(Color online) Recursion procedure combined with corrections to scaling. $(\ell_{\rm eff}/\ell)^d \theta^{n}_{\scriptscriptstyle ++}(0,u_h)$ vs $(\ell_{\rm eff}/\ell)^{(\gamma+\beta)/\nu}u_h$, $n=0$ (red squares), $n=2$ (green dots). Data are for $\ell=9.5$, $\alpha \ell=19.5$, $\delta L_z=1$ and $A=3600$. Corrections to scaling that affect both the amplitude of the function and the reduced parameter $(\ell_{\rm eff}/\ell)^{(\gamma+\beta)/\nu}u_h$ also affects the iteration process so that an effective $\alpha_{\rm eff}=\frac{\alpha \ell + \delta \ell}{\ell + \delta \ell}$ was used.}
\label{iterations_Leff}
\end{figure}

In this appendix we detail the rescaling procedure applied to the function $\theta$ displayed in Fig. \ref{theta-t-h}b).

Figure \ref{before_Leff} shows $\theta^{0}_{\scriptscriptstyle ++}$ obtained with two different sets of system sizes: one was obtained using $\ell=9.5$, $\alpha \ell=19.5$ and $\delta L_z=1$ and the other $\ell=14.5$, $\alpha \ell = 29.5$ and $\delta L_z=1$. The two sets of system sizes give significantly different results which can be attributed in part to corrections to the scaling limit.
Corrections of this amplitude are encountered elsewhere \cite{Ha12,VGMD09}. They  can be accounted for by introducing a phenomenological change to the scaling length \cite{VD13}: $\ell \rightarrow \ell_{\rm eff}=\ell+\delta \ell$, see Fig. \ref{choice_Leff}, a process which can be justified analytically for the Blume-Capel model \cite{toldin_critical_2010}. To obtain a data collapse, we calculate the necessary correction $\delta \ell$ so that $(\ell_{\rm eff}/\ell)^d\theta^{0}_{\scriptscriptstyle ++}$ is equal for the maxima of the two sets of data. We find,  $\delta \ell = 2.8$ with an error of approximately 5 \% considering the statistical error on the data. This correction affects both the amplitude of the function by a factor of $(\ell_{\rm eff}/\ell)^d$ and the reduced parameter $u_h=\tilde{h} L_z^{(\b+\g)/\nu}$ by a factor $(\ell_{\rm eff}/\ell)^{(\gamma+\beta)/\nu}$. We find that this single parameter is enough to make the data collapse both in amplitude and width, as shown in Fig. \ref{choice_Leff}.

When performing iterations following Eq.(\ref{recursion}) on the rescaled data one should use $\alpha_{\rm eff}=\frac{\alpha \ell + \delta \ell}{\ell + \delta \ell}$ rather than $\alpha$. Fig. \ref{iterations_Leff} shows how the approximation $(\ell_{\rm eff}/\ell)^d\theta^{n}_{\scriptscriptstyle ++}$ evolves from $n=0$ to $n=2$, the convergence point of our iteration procedure. 
The function $(\ell_{\rm eff}/\ell)^d\theta^{2}_{\scriptscriptstyle ++}(0,u_h)$ of Fig. \ref{iterations_Leff} obtained using this procedure is in good agreement with that from reference \cite{VD13}, without any further renormalization although our protocol yields a bigger value of $\delta \ell$. Making a best fit between our data and that from Ref. \cite{VD13}, we find a value $\ell_{\rm eff}=2.615$, within 5\% of our independent estimate. This procedure was applied to both data obtained with system sizes $\ell=9.5$, $\alpha \ell=19.5$ and $\ell=14.5$, $\alpha \ell=29.5$, leading to the universal Casimir universal function of Fig. \ref{theta-t-h}. The kink seen in Fig. \ref{theta-t-h-alone} is smoothed out in the rescaling process and the amplitude of the collapsed curves corresponds reasonably to that set by numerical estimates of the universal scaling amplitude, $\Delta_{++}$ (see main text).

\bibliography{biblio_article_2014}

\end{document}